\documentclass[aps,twocolumn,superscriptaddress,showpacs,showkeys]{revtex4}
\usepackage{epsfig}
\usepackage{graphicx}
\input{psfig.sty}

\newcommand{\bastar}{\begin{eqnarray*}}
\newcommand{\eastar}{\end{eqnarray*}}
\newskip\humongous \humongous=0pt plus 1000pt minus 1000pt

\newif\ifdtup

\relax

\newcommand{\be}{\begin{equation}}
\newcommand{\ee}{\end{equation}}
\newcommand{\bea}{\begin{eqnarray}}
\newcommand{\eea}{\end{eqnarray}}
\newcommand{\X}{{\vec X}}
\newcommand{\pro}{\partial}
\newcommand{\n}{\hat n}
\newcommand{\oneg}{\displaystyle\frac{1}{g}}

\newcommand{\D}{{\hat D}}

\newcommand{\valpha}{{\vec \alpha}}

\newcommand{\dfrac}{\displaystyle\frac}
\newcommand{\ba}{\begin{array}}
\newcommand{\ea}{\end{array}}

\newcommand{\nn}{\nonumber}
\newcommand{\mn}{{\mu\nu}}

\newcommand{\Int}{\displaystyle\int}
\begin{document}

\title  {Nucleon Spin in QCD: Old Crisis and New Resolution}
\bigskip

\author{Y. M. Cho}
\email{ymcho@unist.ac.kr}
\affiliation{School of Electrical and Computer Engineering  \\
Ulsan National Institute of Science and Technology, Ulsan 689-798, Korea}
\affiliation{School of Physics and Astronomy,
Seoul National University, Seoul 151-747, Korea}
\author{Mo-Lin Ge}
\affiliation{Theoretical Physics Division, Chern Institute of Mathematics \\
Nankai University, Tianjin 300071, China}
\author{Pengming Zhang}
\affiliation{Institute of Modern Physics \\
Chinese Academy of Science, Lanzhou 730000, China} 

\begin{abstract}
\noindent
\textbf{Abstract:} 
~We discuss the shortfalls of existing resolutions of the long-standing 
gauge invariance problem of the canonical decomposition of the 
nucleon spin to the spin and angular momentum of quarks and gluons. We 
provide two logically flawless expressions of nucleon spin which have 
different physical meanings, using the gauge independent Abelian decomposition. 
The first one is based on the assumption that all gluons (binding and 
valence gluons) contribute to the nucleon spin, but the second one is based on 
the assumption that only the binding gluons (and the quarks) contribute to it. 
We propose the second expression to be the physically correct one.
\end{abstract}
\vspace{0.3cm}
\maketitle
                           
An outstanding problem in nuclear physics is the so-called spin crisis 
problem \cite{bass,stra}. The problem originated from the European Muon 
Collaboration (EMC) experiment on muon scattering from polarized proton, 
which indicated that the contribution of quark spin to proton spin was 
much smaller than theoretical expectation \cite{emc}. As a composite particle 
proton spin should be made of the spin and orbital angular momentum of its 
constituents, and this experimental result was difficult to understand.

To settle this problem we clearly need more precise mesurements on the spin 
and angular momentum of composite particles. The Common Muon Proton Apparatus 
for Structure and Spectroscopy (COMPASS) and similar experiments are expected 
to provide better measurements \cite{exp}. 

Independent of the experimental mesurements, however, 
we face a deeper theoretical question to settle this problem. To compare 
experiment with theory we have to know how to decompose the nucleon spin to 
the spin and angular momentum of the constituents \cite{jaffe,ji}. 
But in gauge theories it is a non-trivial matter to obtain a gauge invariant 
decomposition of the spin of composite particles to the spin and angular momentum 
of the constituents. In fact it has long been suggested that 
this is impossible in gauge theories \cite{ji,jauch}. Nevertheless experiments 
have been able to measure the spin and angular momentum of the constituents 
separately \cite{emc,exp,beth,mar}. This is disturbing. 
{\it The purpose of this Letter is to discuss the present status of 
this problem and to provide the gauge invariant decomposition 
of the spin and angular momentum of quarks and gluons in nucleons.}

To understand the problem, consider the canonical expression of 
the conserved angular momentum in QCD obtained by Noether's theorem,
\bea
&J_\mn^{(qcd)}= S_\mn^q+ L_\mn^q+ S_\mn^g+ L_\mn^g \nn\\
&= \Int {\bar \psi} \gamma^0 \dfrac{\Sigma_\mn}{2} \psi d^3 x
-i \Int {\bar \psi} \gamma^0  x_{[\mu} \pro_{\nu]} \psi d^3 x \nn\\
&- \Int \vec A_{[\mu} \cdot \vec F_{\nu] 0} d^3 x
- \Int \vec F_{0 \alpha} \cdot x_{[\mu} \pro_{\nu]} \vec A_{\alpha} d^3 x.
\label{ns0}
\eea
To be physically meaningful the decomposition has to be gauge independent. 
But three terms except for the first are not gauge invariant, although 
the sum is. This makes the physical meaning of the canonical decomposition 
controversial.

Recently, however, there have been important progresses on this problem \cite{ji,chen,chen2}. 
To see this consider the same problem in atoms, with similar (gauge dependent) 
canonical decomposition $J_\mn^{(qed)}$ derived from QED
\bea  
&J_\mn^{(qed)}=S_\mn^e+ L_\mn^e+ S_\mn^{\gamma}+ L_\mn^{\gamma} \nn\\
&= \Int {\bar \psi} \gamma^0 \dfrac{\Sigma_\mn}{2} \psi d^3 x
-i \Int {\bar \psi} \gamma^0  x_{[\mu} \pro_{\nu]} \psi d^3 x \nn\\
&- \Int A_{[\mu} F_{\nu] 0} d^3 x
- \Int F_{0 \alpha} x_{[\mu} \pro_{\nu]} A_{\alpha} d^3 x. 
\label{asd}
\eea
A best way to make the decomposition gauge invariant is to 
decompose the potential $A_\mu$ first to the vacuum (pure gauge) part 
$\Omega_\mu$ and the physical (transverse) part $X_\mu$,
\bea
&A_\mu=\Omega_\mu + X_\mu,  \nn\\
&\Omega_\mu=\pro_\mu \theta,~~~~~\pro_\mu X_\mu=0.
\label{ad0}
\eea
Notice that the decomposition (\ref{ad0}) is gauge independent because, 
under the gauge transformation we have
\bea
\delta \Omega_\mu= \pro_\mu \alpha,
~~~~~\delta X_\mu=0,
\eea
where $\alpha$ is the gauge parameter. This is due to the fact that 
the connection space (the space of potentials) forms an affine space. 
Now, adding a surface term to (\ref{asd}) one can modify it to 
\bea
&J_\mn^{(qed)}= \Int {\bar \psi} \gamma^0 \dfrac{\Sigma_\mn}{2} \psi d^3 x
-i \Int {\bar \psi} \gamma^0  x_{[\mu} {\bar D}_{\nu]} \psi d^3 x \nn\\
&-\Int X_{[\mu} F_{\nu] 0} d^3 x
- \Int F_{0 \alpha} x_{[\mu} \pro_{\nu]} X_{\alpha} d^3 x,
\label{qed}
\eea
where $\bar D_\mu=\pro_\mu-ie\Omega_\mu$. Clearly each term in this 
expression is gauge invariant. Moreover this perfectly accounts for 
the experimental results which mesure each term separately \cite{beth,mar}. 
This shows that one can indeed express the spin of composite particles 
with those of the constituents in a self-consistent and gauge invariant way. 

One could obtain similar gauge invariant expression for nucleon spin 
decomposing the non-Abelian gauge potential to the vacuum and physical parts
$\hat \Omega_\mu$ and $\vec Z_\mu$ which satisfy the desired gauge transformation 
property 
\bea
&\vec A_\mu=\hat \Omega_\mu + \vec Z_\mu, \nn\\
&\delta \hat \Omega_\mu= \dfrac{1}{g} \bar D_\mu \vec \alpha,
~~~~~\delta \vec Z_\mu= -\vec \alpha \times \vec Z_\mu,
\label{vdec}
\eea
where $\bar D_\mu=\pro_\mu+g \hat \Omega \times$ and $\vec \alpha$ is the 
(infinitesimal) gauge parameter. The difficult question 
here is how can one make such decomposition.  

It has been proposed that one can make such decomposition requiring \cite{chen} 
\bea
&\bar D_{[\mu} \hat \Omega_{\nu]}=0,~~~~~\vec Z_\mu \times \vec F_{\mu 0}=0.
\label{cc}
\eea
Moreover, with this the following gauge invariant decomposition of 
the nucleon spin has been proposed
\bea
&J_\mn^{(qcd)'}= \Int {\bar \psi} \gamma^0 \dfrac{\Sigma_\mn}{2} \psi d^3 x
-i \Int {\bar \psi} \gamma^0  x_{[\mu} {\bar D}_{\nu]} \psi d^3 x \nn\\
&- \Int \vec Z_{[\mu} \cdot \vec F_{\nu] 0} d^3 x
- \Int \vec F_{0 \alpha} \cdot x_{[\mu} \pro_{\nu]} \vec Z_{\alpha} d^3 x, 
\label{ns1}
\eea
The justification for this was that $\hat \Omega_\mu$ and $\vec Z_\mu$ 
so defined make each term in (\ref{ns1}) satisfy angular momentum algebra
and at the same time gauge invariant. Moreover, $\hat \Omega_\mu$ and 
$\vec Z_\mu$ which satisfy (\ref{cc}) remain so after the gauge 
transformation \cite{chen}.

This proposal is interesting, but has potentially 
serious problems \cite{ji2,waka}. For example, it is not clear that 
(\ref{cc}) can uniquely determine $\hat \Omega_\mu$ and $\vec Z_\mu$
and provide the desired decomposition. 
It does not assure that $\hat \Omega_\mu$ represents the vacuum,
nor does it guarantee the necessary requirement that $\vec Z_\mu$ is 
transverse. Moreover, even if (\ref{cc}) does so, it is very difficult 
to express them in a physically meaningful way. So it is very hard to 
figure out what experiments can measure different terms in (\ref{ns1}). 

To cure these defects an improvement of (\ref{ns1}) which replaces the plain 
derivative of the last term by $\bar D_\mu$ with $\bar D_\mu \vec Z_\mu=0$ 
has been proposed \cite{chen2}. But this improvement still has a critical 
defect in that it can not identify the vacuum potential, which one need 
to make the decomposition (\ref{vdec}).    

Actually there is a well-known gauge independent decomposition of the non-Abelian 
gauge potential to the vacuum and physical parts \cite{prd80,prl81}.
Consider SU(2) QCD for simplicity, and let $\n_i~(i=1,2,3)$ be a gauge covariant 
right-handed orthonormal basis in SU(2) space. Then we can define the most 
general vacuum imposing the maximal magnetic symmetry to the potential $\vec A_\mu$,
\bea
&\forall_i~~~D_\mu \n_i=0.~~~~~(\hat n_i^2=1)
\label{vp}
\eea
Indeed from this we have $[D_\mu, D_\nu]~\n_i=g \vec F_\mn \times \n_i=0$, which assures
$\vec F_\mn=0$. Solving (\ref{vp}) we obtain the most general vacuum potential \cite{plb07}
\bea
\hat \Omega_\mu= \dfrac{1}{2g} \epsilon_{ijk} (\n_i \cdot \pro_\mu \n_j)~\n_k.
\label{vps}
\eea
Moreover, we can easily prove that (\ref{vdec}) with (\ref{vps}) provides 
a gauge independent decomposition which has the desired gauge transformation 
property (with $\delta \n_i=-\vec \alpha \times \n_i$). Again this 
is because $\hat \Omega_\mu$ defined by (\ref{vp}) forms a connection space. 
Moreover, we can make $\vec Z_\mu$ physical requiring the transversality 
condition \cite{prd00}
\bea
&\bar D_\mu \vec Z_\mu=0,~~~~~(\bar D_\mu= \pro_\mu +g \hat \Omega_\mu \times).
\label{tc}
\eea
Obviously this is the generalization of (\ref{ad0}) to QCD which provides the desired 
decomposition.

With this we may have
\bea
&J_\mn^{(qcd)}= \Int {\bar \psi} \gamma^0 \dfrac{\Sigma_\mn}{2} \psi d^3 x
-i \Int {\bar \psi} \gamma^0  x_{[\mu} {\bar D}_{\nu]} \psi d^3 x \nn\\
&- \Int \vec Z_{[\mu} \cdot \vec F_{\nu] 0} d^3 x
- \Int \vec F_{0 \alpha} \cdot x_{[\mu} \bar D_{\nu]} \vec Z_{\alpha} d^3 x.
\label{ns2}
\eea
This is equivalent to the canonical expression (\ref{ns0}), up to the surface term
\bea
\int \partial_\alpha \big( \vec F_{0\alpha} \cdot x_{[\mu} \hat \Omega_{\nu]} \big) d^3 x.
\eea
Moreover, each term here is explicitly gauge invariant and at the same time satisfies 
the angular momentum algebra. So this is the gauge invariant modification of (\ref{ns0})
which provides the correct generalization of (\ref{qed}) to QCD. Clearly 
(\ref{ns2}) is formally identical to the expression proposed as an improvement 
of (\ref{ns1}) \cite{chen2}. But there is a big difference. Here we now have (\ref{vps}),
which tells exactly what is the vacuum potential. 
 
Nevertheless we strongly doubt that this can correctly describe the nucleon 
spin. To see this notice that there are actually two types of gluons, the 
binding gluons and the valence gluons, and two types of QCD, the standard 
QCD and the restricted QCD which we call RCD \cite{prd80,prl81}. And QCD 
becomes RCD made of the binding gluons which has the valence gluons as 
the colored source. This means that the valence gluons (just like the quarks) 
become colored source which have to be confined, so that they play no role 
in confinement. So the confinement should come only through the binding gluons. 
This is known as the Abelian dominance in QCD \cite{prd80,prl81,prd00,thooft}. 

Most importantly, the quark model of hadrons tells that (low-lying) nucleons 
are made of three quarks (and binding gluons), but not valence gluons (color 
singlets made of $3\times 3\times 3$, not $3\times 3\times 3\times 8$) \cite{pdg}. 
In other words valence gluons are not regarded as the constituents of nucleons. 
Instead they become the constituents of glueballs. If so, only quarks and binding 
gluons should contribute to the nucleon spin. But this important point has 
completely been ignored in the above discussion. 

To exclude the valence gluons in (\ref{ns2}) we have to 
separate the valence gluons from the binding gluons. This can be done by 
the Abelian decomposition \cite{prd80,prl81}. Let $\n=\n_3$ be the unit vector 
which selects the color direction at each space-time point, and make 
the Abelian projection imposing only one magnetic symmetry to $\vec A_\mu$,
\bea
D_\mu \n = \pro_\mu \n + g {\vec A}_\mu \times \n = 0.~~~(\n^2=1)
\label{ap}
\eea
This selects the Abelian part of the potential (the restricted potential)
\bea
&\hat A_\mu =A_\mu \n - \oneg \n \times\pro_\mu \n.
~~~(A_\mu = \n\cdot \vec A_\mu)
\label{rp}
\eea
where $A_\mu$ is the ``electric'' potential. With this we have the Abelian 
decomposition of the non-Abelian gauge potential \cite{prd80,prl81},
\bea
& \vec{A}_\mu = \hat A_\mu + \X_\mu.~~~(\hat{n} \cdot \vec{X}_\mu=0).
\label{adec}
\eea
What is important about this decomposition is that it is gauge independent. 
Once $\n$ is chosen, the decomposition follows automatically, independent 
of the choice of a gauge. 

The restricted potential $\hat A_\mu$ by itself forms a connection 
space, so that under the (infinitesimal) gauge transformation we have
\bea
&\delta \hat A_\mu = \oneg \D_\mu \valpha,
~~~~~\delta \X_\mu = - \valpha \times \X_\mu.
\eea
For this reason $\hat A_\mu$ and $\vec X_\mu$ are called the binding 
potential and the valence potential. Moerover, $\hat A_\mu$ has a dual 
structure \cite{prd80,prl81}
\begin{eqnarray}
& \hat{F}_{\mu\nu} = (F_{\mu\nu}+ H_{\mu\nu})\hat{n}\mbox{,}\nonumber \\
& F_{\mu\nu} = \partial_\mu A_{\nu}-\partial_{\nu}A_\mu \mbox{,}\nonumber \\
& H_{\mu\nu} = -\dfrac{1}{g} \hat{n}\cdot(\partial_\mu
\hat{n}\times\partial_\nu\hat{n})
= \partial_\mu \tilde C_\nu-\partial_\nu \tilde C_\mu,
\label{rfs}
\end{eqnarray}
where $\tilde C_\mu$ is the ``magnetic'' potential which 
represents the non-Abelian monopole originating from the last term of 
(\ref{rp}). Moreover, with (\ref{adec}) we can easily show that 
QCD can be viewed as RCD made of (\ref{rfs}) which has the valence gluons 
as its source \cite{prd80,prl81}. 

Since $\hat A_\mu$ still contains the (unphysical) pure 
gauge degrees of freedom, we need to separate the physical part. 
This can be done decomposing it further to the vacuum  
and physical parts, 
\bea
&\hat A_\mu= \hat \Omega_\mu+ \vec B_\mu,
~~~~~\bar D_\mu \vec B_\mu=0,   \nn\\
&\vec B_\mu= B_\mu \n,~~~~B_\mu= A_\mu- \dfrac{1}{g} \n_1 \cdot \pro_\mu \n_2.
\label{adec2}
\eea
Notice that $\vec B_\mu$ (just like $\vec X_\mu$) transforms covariantly 
under the gauge transformation. This is because $\hat A_\mu$ and $\hat \Omega_\mu$ 
themself form connection space. This tells that $B_\mu$ (like $X_\mu$ 
in QED) is gauge invariant. Moreover, the transversality condition
for $\vec B_\mu$ assures $\pro_\mu B_\mu=0$. 

Now, it is straightforward to obtain the desired expression of nucleon spin. 
All we have to do is to replace $\vec A_\mu$ and $\vec F_\mn$ by 
$\hat A_\mu$ and $\hat F_\mn$ in (\ref{ns2}). So we finally have
\bea
&J_\mn^{(qcd)}= \Int {\bar \psi} \gamma^0 \dfrac{\Sigma_\mn}{2} \psi d^3 x
-i \Int {\bar \psi} \gamma^0  x_{[\mu} {\bar D}_{\nu]} \psi d^3 x \nn\\
&- \Int \vec B_{[\mu} \cdot \hat F_{\nu] 0} d^3 x
- \Int \hat F_{0 \alpha} \cdot x_{[\mu} {\bar D}_{\nu]} \vec B_{\alpha} d^3 x,
\label{ns3}
\eea 
where $\vec B_\mu$ is the transverse binding gluon. Clearly each term is 
gauge invariant, and can be shown to satisfy the angular momentum algebra 
separately. 

One may ask what is the observable difference between (\ref{ns2}) and 
(\ref{ns3}). An obvious difference is that (\ref{ns3}) has no contribution 
from valence gluons. Another difference is that (\ref{ns3}) is almost identical 
to QED expression (\ref{qed}). This is almost evident, but to see this more 
clearly remember that we can always have the gauge independent Abelianization 
of QCD. In this Abelianization, QCD becomes an Abelian gauge theory with two 
(electric and magnetic) U(1) potentials coupled to valence gluons \cite{prd00,prd02}. 
So, by Abelianizing (\ref{ns3}) we can indeed make it almost identical to (\ref{qed}). 
This observation can be very useful to compare (\ref{ns3}) with experiment.
  
At this point one may wonder if there is any theory which can justify (\ref{ns3}), 
and if so, what is such theory. There is, of course, and the underlying theory is  
RCD \cite{prd80,prl81}. In fact from RCD we obtain the following canonical (Noether's)
expression of the conserved nucleon spin, 
\bea
&J_\mn^{(rcd)}= \Int {\bar \psi} \gamma^0 \dfrac{\Sigma_\mn}{2} \psi d^3 x
-i \Int {\bar \psi} \gamma^0  x_{[\mu} \pro_{\nu]} \psi d^3 x \nn\\
&- \Int \hat A_{[\mu} \cdot \hat F_{\nu] 0} d^3 x
- \Int \hat F_{0 \alpha} \cdot x_{[\mu} \pro_{\nu]} \hat A_{\alpha} d^3 x.
\label{nsrcd}
\eea
Moreover, we can easily show that this is equivalent to (\ref{ns3}), up to 
a surface term. In this sense the expression $J_\mn^{(qcd)}$ in (\ref{ns3}) 
should more properly be written as $J_\mn^{(rcd)}$.

In this paper we have kept the Lorentz covariance intact, so that all our 
expressions have Lorentz degrees of freedom. So in practical applications 
one can simplify our expressions further choosing the prefered Lorentz frame.

To summarize, we have presented two gauge invariant decompositions (\ref{ns2}) 
and (\ref{ns3}) of nucleon spin, and argued that (\ref{ns3}) is the correct one. 
But ultimately experiments should determine which is correct. If (\ref{ns3}) 
is endorsed by experiments, it would merely reconfirm the quark model of nucleons. 
But if by any means (\ref{ns2}) is endorsed, the quark model 
will be seriously challenged.

Our argument is based on the gauge independent decomposition of non-Abelian 
gauge potential. We have discussed three decompositions, the decomposition 
of the potential to the restricted and valence parts,
the decomposition of the restricted potential to the vacuum 
and binding parts, and the decomposition of the potential
to the vacuum and physical parts. All are based on 
the fact that the connection space forms an affine space.

Perhaps a more important issue closely related to the above nucleon spin 
is the decomposition of nucleon momentum, which has become an hot 
issue \cite{chen2,ji2,waka}. Clearly our analysis should have deep 
impact on this issue, because it implies that the valence gluons 
should again have no place in nucleon momentum. We hope to discuss this 
issue in more detail in a separate paper.   

We conclude with the following remarks: \\
1) The Abelian decomposition has played a crucial role to prove the 
Abelian dominance and color confinement in QCD \cite{prd80,prl81}. 
In particular, it played the pivotal role to resolve the problem of 
the gauge dependence of the monopole condensation in QCD \cite{prd02,kondo}.  
Here we show that the further decomposition of the restricted potential 
to the binding and vacuum potentials plays important role 
in analyzing the nucleon spin. \nn\\ 
2) Our discussion tells that RCD can play important roles in nuclear physics. 
It has the full non-Abelian gauge symmetry and all ingredients to prove 
the confinement \cite{prd00,prd02}. So it has everything to describe 
the nucleons, and recent lattice calculations support this \cite{pdg,kondo}.  \nn\\ 
3) In this paper we have considered SU(2) QCD for simplicity. But 
in reality QCD is based on SU(3) so that we have to generalize our discussion 
to SU(3). This is non-trivial, but straightforward. And all our conclusions 
hold in SU(3) QCD.  

A detailed discussion and the generalization of our result to SU(3) QCD will be
presented elsewhere \cite{cho}.

{\bf ACKNOWLEDGEMENT}

The work is supported in part by National Research Foundation 
(Grant 2010-002-1564) of Korea and by National Natural Science Foundation 
(Grants 10604024, 11035006, and 11075077) of China.

\end{document}